\documentclass[pra,twocolumn,superscriptaddress,showpacs]{revtex4}
\newcommand{\UQ}{ARC Centre of Excellence for Quantum-Atom Optics, 
School of Mathematics and Physics, University of Queensland, QLD 4072, Australia.}
\newcommand{\NZ}{Jack Dodd Centre for Quantum Technology, Department of Physics, University of Otago, P. O. Box 56, Dunedin, New Zealand}
\newcommand{\VIR}{Department of Physics, University of Virginia, 382 McCormick Road, Charlottesville, Virginia 22904-4714, USA}
\usepackage{graphicx,psfrag}
\usepackage{bm}
\usepackage{amssymb}
\usepackage{graphicx}
\usepackage{bm}
\usepackage{color}
\usepackage{psfrag}
\usepackage{epstopdf}
\usepackage{amsfonts}
\usepackage{amsmath}
\usepackage{amsbsy} 

\newcommand{\pr}{Phys. Rev. }
\newcommand{\jpb}{J. Phys. B }

\newcommand{\jpa}{J. Phys. A }

\newcommand{\josas}{J. Opt. Soc. Am.}

\begin{document}
\title{Quadripartite continuous-variable entanglement via quadruply concurrent downconversion}

\author{S.~L.~W. Midgley}
\affiliation{\UQ}
\author{A.~S. Bradley}
\affiliation{\NZ}
\author{O. Pfister}
\affiliation{\VIR}
\author{M.~K. Olsen}
\affiliation{\UQ}

\date{\today}

\begin{abstract}

We investigate an intra-cavity coupled down-conversion scheme to generate quadripartite entanglement using concurrently resonant nonlinearities. We verify that quadripartite entanglement is present in this system by calculating the output fluctuation spectra and then considering violations of optimized inequalities of the van Loock-Furusawa type. The entanglement characteristics both above and below the oscillation threshold are considered. We also present analytic solutions for the quadrature operators and the van Loock-Furusawa correlations in the undepleted pump approximation.\end{abstract}


\maketitle

\section{Introduction}

Entanglement is a concept of central importance in quantum theory and continues to inspire both theoretical and experimental efforts to explore quantum systems. In addition to this, entanglement is the main resource of quantum information and in particular, multipartite continuous-variable (CV) entangled states have grown to be pivotal in multipartite quantum communication \cite{vanny,vanny1,vanny2,jing,aoki,yone,zhang}. The criteria which must be satisfied to establish whether bipartite entanglement exists in a given system are well known for the CV case \cite{Duan, Simon}. Furthermore, bipartite entanglement can be realised experimentally. The criteria for the bipartite scenario have been generalized for multipartite entanglement by van Loock and Furusawa \cite{furu}. Advances have also taken place in the experimental generation of tripartite entanglement. In particular, there have been experiments where entangled beams are produced by mixing squeezed beams with linear optics \cite{vanny,jing,afuru,aoki, yone, silb}. Several proposals have also been made whereby multi-frequency entangled outputs are generated. These rely on the use of non-degenerate downconversion \cite{coelho} or cascaded or concurrent nonlinear optical processes \cite{gao, ferraro, mko3, muz, mko2} where the tripartite entanglement is instead produced via the interaction with the nonlinear medium itself. It is these concurrent processes that we consider in this work for the quadripartite case.

In regards to quadripartite entangled beams there have been theoretical proposals based on linear optics and cascaded nonlinearites \cite{su, wang1, wang2, leng}. In this work we build on a tripartite scheme proposed by Bradley et al. in \cite{mko3} but for the case of quadruply concurrent nonlinearities. Furthermore, we use an optimized version of the van Loock-Furusawa inequalities to demonstrate quadripartite entanglement in this system.

This paper is organized as follows. In Section II we describe the Hamiltonian for the system under consideration. Section III 
discusses the van Loock-Furusawa (VLF) criteria as a means of quantifying quadripartite entanglement. Section IV considers the interaction Hamiltonian in the undepleted pump approximation and gives analytic solutions for the quadrature operators, as well as the VLF correlations. In Section V we present the full equations of motion for the system and Section VI gives the steady state solutions to the classical versions of the equations of motion and provides an overview of the linearized fluctuation analysis used in this work to calculate the measurable output fluctuation spectra from the cavity. These output spectra are are found in Section VII and used to demonstrate violation of the optimized van Loock-Furusawa criteria and hence, demonstrate quadripartite entanglement.

\section{System and Hamiltonian}

\begin{figure}[htb!]
\includegraphics[width=0.5\textwidth]{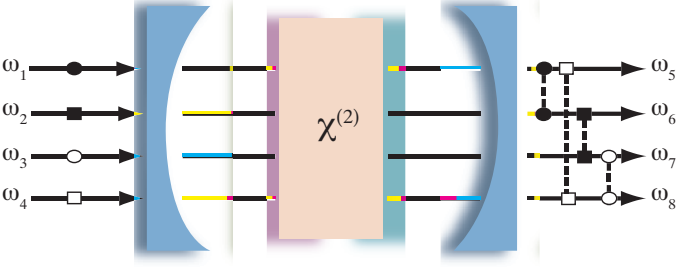}
\caption{(Color online) Schematic of a $\chi^{(2)}$ crystal inside a pumped Fabry-P\'{e}rot cavity. Pump lasers drive four intracavity modes with frequencies $\omega_{1}$, $\omega_{2}$, $\omega_{3}$ and $\omega_{4}$ (represented by circles and squares), which are down-converted to four output modes with frequencies $\omega_{5}$, $\omega_{6}$, $\omega_{7}$ and $\omega_{8}$.}
\label{fig:fig2}
\end{figure}

We model a system in which pump lasers drive four modes in an optical cavity. As depicted in the simplified experimental setup shown in Figure~\ref{fig:fig2}, the four inputs interact with a $\chi^{(2)}$ non-linear crystal to produce four low-frequency entangled output beams at frequencies $\omega_{5},\omega_{6},\omega_{7},\omega_{8}$. For example, mode 1 is pumped at a particular frequency and polarization such that it produces modes 5 and 6.

The full Hamiltonian for the eight-mode system, describing the interaction inside the optical cavity and the interaction of the cavity fields with the output fields, can be written as
\begin{equation}
{\cal H} = {\cal H}_{pump}+{\cal H}_{int}+{\cal H}_{free}+{\cal H}_{res},
\label{eq:Hfull}
\end{equation}

\noindent where the interaction Hamiltonian is
\begin{eqnarray}
{\cal H}_{int} &=& i\hbar[\chi_{1}\hat{a}_{1}\hat{a}_{5}^{\dagger}\hat{a}_{6}^{\dagger} + \chi_{2}\hat{a}_{2}\hat{a}_{6}^{\dagger}\hat{a}_{7}^{\dagger} +\chi_{3}\hat{a}_{3}\hat{a}_{7}^{\dagger}\hat{a}_{8}^{\dagger}\nonumber\\
&+&\chi_{4}\hat{a}_{4}\hat{a}_{8}^{\dagger}\hat{a}_{5}^{\dagger}] + \textnormal{H.c.},
\label{eq:Hint}
\end{eqnarray}

\noindent with the $\chi_{i}$ representing the effective nonlinearities and $\hat{a}_{i}$ denoting the bosonic annihilation operators for the intracavity modes at frequencies $\omega_{i}$. The pumping Hamiltonian, describing the the cavity driving fields, in the appropriate rotating frame is

\begin{equation}
{\cal H}_{pump} = i\hbar\sum_{i=1}^{4}\left[\epsilon_{i}\hat{a}_{i}^{\dag}-\epsilon_{i}^{\ast}\hat{a}_{i}\right],
\label{eq:Hpump}
\end{equation}

\noindent and the cavity damping Hamiltonian is given by

\begin{equation}
{\cal H}_{res}= \hbar\sum_{i=1}^{8}\left[\hat{\Gamma}_{i}\hat{a}_{i}^{\dag}+\hat{\Gamma}_{i}^{\dag}\hat{a}_{i}\right],
\label{eq:Hres}
\end{equation}

\noindent where $\epsilon_{i}$ are the classical pumping laser amplitudes for mode $i$, and the $\hat{\Gamma}_{i}$ are the annihilation operators for bath quanta to which each of the intracavity modes are coupled, and which represent losses through the cavity mirror. 

\section{Quadripartite Entanglement Measures}
\label{III}

In order to investigate multipartite entanglement, and in particular show that the system under consideration demonstrates true quadripartite entanglement, we first define quadrature operators \cite{reid} for each mode as,

\begin{equation}
\hat{X}_{i}=\hat{a}_{i}+\hat{a}^{\dagger}_{i}, \hspace{0.2cm} \hat{Y}_{i}=-i(\hat{a}_{i}-\hat{a}^{\dagger}_{i})
\label{eqn20}
\end{equation}

\noindent such that $[\hat{X}_{i},\hat{Y}_{i}]=2i$. Based on this definition $V(\hat{X}_{i}) \le 1$, for example, indicates single-mode squeezing where $V(\hat{A})=\langle\hat{A}^{2}\rangle-\langle\hat{A}\rangle^{2}$ denotes the variance. The conditions proposed by van Loock and Furusawa \cite{furu}, which are a generalization of the conditions for bipartite entanglement \cite{Duan, Simon} are sufficient to demonstrate multipartite entanglement. We now demonstrate how these may be optimized for the verification of genuine quadripartite entanglement in this system. 

Using the quadrature definitions in Equation~(\ref{eqn20}), the quadripartite inequalities which must be simultaneously violated by the low frequency modes are,

\begin{eqnarray}\label{V56}
V(\hat{X}_{5}-\hat{X}_{6}) + V(\hat{Y}_{5} + \hat{Y}_{6} + g_{7}\hat{Y}_{7} + g_{8}\hat{Y}_{8}) \ge 4,\\\label{V67}
V(\hat{X}_{6}-\hat{X}_{7}) + V(g_{5}\hat{Y}_{5} + \hat{Y}_{6} + \hat{Y}_{7} + g_{8}\hat{Y}_{8}) \ge 4,\\\label{V78}
V(\hat{X}_{7}-\hat{X}_{8}) + V(g_{5}\hat{Y}_{5} + g_{6}\hat{Y}_{6} + \hat{Y}_{7} + \hat{Y}_{8}) \ge 4,
\end{eqnarray}

\noindent where the $g_{i}  (i=5,6,7,8)$ are arbitrary real parameters that are used to optimize the violation of these inequalities. In particular, we minimize Equations (\ref{V56}) and (\ref{V78}) with respect to $g_{7,8}$ and $g_{5,6}$, respectively. Solving the resulting equations leads to the optimized expressions,

\begin{eqnarray}
g_{5} &=& \frac{V_{6}(V_{57}+V_{58})-V_{56}(V_{67}+V_{68})}{V^{2}_{56}-V_{5}V_{6}},\\
g_{6} &=& \frac{V_{5}(V_{67}+V_{68})-V_{56}(V_{57}+V_{58})}{V^{2}_{56}-V_{5}V_{6}},\\
g_{7} &=& \frac{V_{8}(V_{57}+V_{67})-V_{78}(V_{58}+V_{68})}{V^{2}_{78}-V_{7}V_{8}},\\
g_{8} &=& \frac{V_{7}(V_{58}+V_{68})-V_{78}(V_{57}+V_{67})}{V^{2}_{78}-V_{7}V_{8}},
\end{eqnarray}

\noindent where for covariances we use the notation $V_{ij}=(\langle \hat{Y}_{i}\hat{Y}_{j}\rangle+\langle \hat{Y}_{j}\hat{Y}_{i}\rangle)/2-\langle \hat{Y}_{i}\rangle\langle \hat{Y}_{j}\rangle$ and for the case where $i=j$ the covariance, denoted $V_{i}$, reduces to the usual variance, $V(\hat{Y}_{i})$. It is important to note that in the uncorrelated limit these optimized van Loock-Furusawa criteria approach 4. Hence, without optimization some entanglement which is present may be missed.

\section{Analytic solutions in the undepleted pump approximation}

It is useful to consider the interaction Hamiltonian in the undepleted pump approximation in the absence of a cavity in advance of a more complete approach that considers the full quantum equations of motion for all of the interacting fields inside a cavity. We stress here that these equations are not of exact physical relevance, but do give useful insights into the properties of the Hamiltonian. Here we show that it is possible to obtain analytic solutions for the quadrature operator equations of motion using the undepleted pump approximation. This entails setting  $\xi_{i}=\chi_{i}\langle \hat{a}_{i}(0)\rangle$ ($i$=1,2,3,4), where $\xi_{i}$ are positive, real constants. Under these conditions the interaction Hamiltonian can be written as,

\begin{eqnarray}
{\cal H}_{int} &=& i\hbar\Big[\xi_{1}(\hat{a}_{5}^{\dagger}\hat{a}_{6}^{\dagger}-\hat{a}_{5}\hat{a}_{6}) + \xi_{2}(\hat{a}_{6}^{\dagger}\hat{a}_{7}^{\dagger}-\hat{a}_{6}\hat{a}_{7}) \nonumber\\
&+&\xi_{3}(\hat{a}_{7}^{\dagger}\hat{a}_{8}^{\dagger}-\hat{a}_{7}\hat{a}_{8})+\xi_{4}(\hat{a}_{8}^{\dagger}\hat{a}_{5}^{\dagger}-\hat{a}_{8}\hat{a}_{5})\Big].
\label{eq:Hint2}
\end{eqnarray}

The Heisenberg equations of motion can then be written,

\begin{eqnarray}
\frac{d\hat{a}_{5}}{dt}&=&\xi_{1}\hat{a}_{6}^{\dagger}+\xi_{4}\hat{a}_{8}^{\dagger},\\
\frac{d\hat{a}_{6}}{dt}&=&\xi_{1}\hat{a}_{5}^{\dagger}+\xi_{2}\hat{a}_{7}^{\dagger},\\
\frac{d\hat{a}_{7}}{dt}&=&\xi_{2}\hat{a}_{6}^{\dagger}+\xi_{3}\hat{a}_{8}^{\dagger},\\
\frac{d\hat{a}_{8}}{dt}&=&\xi_{3}\hat{a}_{7}^{\dagger}+\xi_{4}\hat{a}_{5}^{\dagger},
\end{eqnarray}

\noindent and these equations can be recast in terms of the quadrature operators as follows,

\begin{eqnarray}
\frac{d\hat{X}_{5}}{dt}&=&\xi_{1}\hat{X}_{6}+\xi_{4}\hat{X}_{8},\\
\frac{d\hat{Y}_{5}}{dt}&=&-\xi_{1}\hat{Y}_{6}-\xi_{4}\hat{Y}_{8},\\
\frac{d\hat{X}_{6}}{dt}&=&\xi_{1}\hat{X}_{5}+\xi_{2}\hat{X}_{7},\\
\frac{d\hat{Y}_{6}}{dt}&=&-\xi_{1}\hat{Y}_{5}-\xi_{2}\hat{Y}_{7},\\
\frac{d\hat{X}_{7}}{dt}&=&\xi_{2}\hat{X}_{6}+\xi_{3}\hat{X}_{8},\\
\frac{d\hat{Y}_{7}}{dt}&=&-\xi_{2}\hat{Y}_{6}-\xi_{3}\hat{Y}_{8},\\
\frac{d\hat{X}_{8}}{dt}&=&\xi_{3}\hat{X}_{7}+\xi_{4}\hat{X}_{5},\\
\frac{d\hat{Y}_{8}}{dt}&=&-\xi_{3}\hat{Y}_{7}-\xi_{4}\hat{Y}_{5}.
\end{eqnarray}

\noindent It is these equations that we solve to find analytic solutions for the quadrature operators as functions of their initial values. 

\subsection{Solutions with equal $\xi_{i}$ }
\label {caseA}

To begin with, we set all the interactions equal so that $\xi_{i}=\xi$ and find analytic expressions for the VLF correlations by solving the Heisenberg equations of motion for this case. The solutions for the quadrature operators are found to be,

\begin{eqnarray}
\hat{X}_{5}(t)&=&A\hat{X}_{5}(0)+B\hat{X}_{6}(0)+C\hat{X}_{7}(0)+B\hat{X}_{8}(0),\\
\hat{Y}_{5}(t)&=&A\hat{Y}_{5}(0)-B\hat{Y}_{6}(0)+C\hat{Y}_{7}(0)-B\hat{Y}_{8}(0),\\
\hat{X}_{6}(t)&=&B\hat{X}_{5}(0)+A\hat{X}_{6}(0)+B\hat{X}_{7}(0)+C\hat{X}_{8}(0),\\
\hat{Y}_{6}(t)&=&-B\hat{Y}_{5}(0)+A\hat{Y}_{6}(0)-B\hat{Y}_{7}(0)+C\hat{Y}_{8}(0),\\
\hat{X}_{7}(t)&=&C\hat{X}_{5}(0)+B\hat{X}_{6}(0)+A\hat{X}_{7}(0)+B\hat{X}_{8}(0),\\
\hat{Y}_{7}(t)&=&C\hat{Y}_{5}(0)-B\hat{Y}_{6}(0)+A\hat{Y}_{7}(0)-B\hat{Y}_{8}(0),\\
\hat{X}_{8}(t)&=&B\hat{X}_{5}(0)+C\hat{X}_{6}(0)+B\hat{X}_{7}(0)+A\hat{X}_{8}(0),\\
\hat{Y}_{8}(t)&=&-B\hat{Y}_{5}(0)+C\hat{Y}_{6}(0)-B\hat{Y}_{7}(0)+A\hat{Y}_{8}(0),
\end{eqnarray}

\noindent where

\begin{eqnarray}
A&=&\cosh^{2}(\xi t),\\
B&=&\frac{1}{2}\sinh(2\xi t),\\
C&=&\sinh^{2}(\xi t).
\end{eqnarray}

From these expressions for the quadrature operators it is possible to find the variances and covariances necessary to calculate the van Loock-Furusawa (VLF) criteria within the undepleted pump approximation. In fact, the variances are all equal and given by the following time-dependent moments,

\begin{equation}
\langle\hat{X}_{i}^{2}\rangle=\langle\hat{Y}_{i}^{2}\rangle=A^{2}+2B^{2}+C^{2},
\end{equation}

\noindent since the expectation values of the amplitudes are all zero. Here we have used the fact that $\langle\hat{X}_{i}(0)\hat{X}_{j}(0)\rangle=\langle\hat{Y}_{i}(0)\hat{Y}_{j}(0)\rangle=\delta_{ij}$. A similar approach can be used to calculate the covariances which are equivalent to the time-dependent moments $\langle\hat{X}_{i}\hat{X}_{j}\rangle$ and $\langle\hat{Y}_{i}\hat{Y}_{j}\rangle$. In particular, the covariances are given by,

\begin{eqnarray}
\langle\hat{X}_{5}{X}_{6}\rangle&=&-\langle\hat{Y}_{5}{Y}_{6}\rangle=2(AB+BC),\\
\langle\hat{X}_{5}{X}_{7}\rangle&=&\langle\hat{Y}_{5}{Y}_{7}\rangle=2(AC+B^{2}),\\
\langle\hat{X}_{5}{X}_{8}\rangle&=&-\langle\hat{Y}_{5}{Y}_{8}\rangle=2(AB+BC),\\
\langle\hat{X}_{6}{X}_{7}\rangle&=&-\langle\hat{Y}_{6}{Y}_{7}\rangle=2(AB+BC),\\
\langle\hat{X}_{6}{X}_{8}\rangle&=&\langle\hat{Y}_{6}{Y}_{8}\rangle=2(AC+B^{2}),\\
\langle\hat{X}_{7}{X}_{8}\rangle&=&-\langle\hat{Y}_{7}{Y}_{8}\rangle=2(AB+BC).
\end{eqnarray}

\begin{figure}[ht!]
\includegraphics[width=0.85\columnwidth]{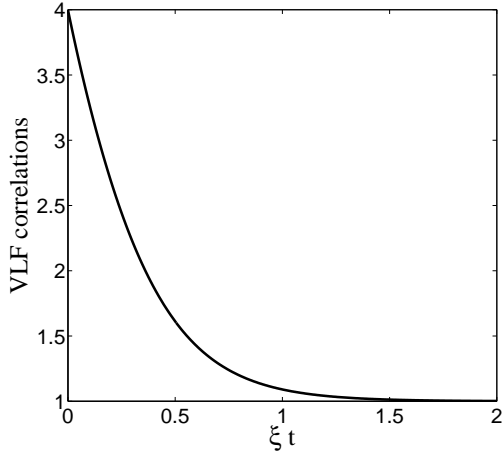} 
\caption{Analytic solutions for the optimized van Loock-Furusawa correlations, $V_{3}$, found by solving the Heisenberg equations of motion in the undepleted pump approximation. A value of less than 4 signifies quadripartite entanglement. All quantities depicted here, and in subsequent graphs, are dimensionless. }
\label{fig0}
\end{figure}

\noindent From these variances and covariances we obtain Equation~\ref{eq68}, an analytic expression for the optimized VLF correlations defined in Equations~\ref{V56}-\ref{V78}. All three VLF correlations are equal when $\xi_{i}=\xi$ and hence we label any of the correlations in Equations~\ref{V56}-\ref{V78} as $V_{3}$. Figure~\ref{fig0} provides a plot of these optimized VLF correlations, $V_{3}$, as a function of $\xi t$. We observe that quadripartite entanglement is present in this system with $V_{3}<4$ for all $\xi t$. This suggests a more complete treatment incorporating depletion of the pump fields and a cavity, will find where quadripartite entanglement is present.

\begin{widetext}
\begin{eqnarray}
V_{3}=4A^{2}-4A\sqrt{2B}+4(\sqrt{2B}-C)C+\frac{2(B^{2}-4B^{3/2}C\sqrt{2}+12BC^{2}-8C^{3}\sqrt{2B}+4C^{4})}{A^{2}-2A\sqrt{B}+B-C\sqrt{2B}+C^{2}}
\label{eq68}
\end{eqnarray}
\end{widetext}

\subsection{Solutions with unequal $\xi_{i}$  }

For simplicity, here we assume that $\xi_{1}=\xi_{2}$ and $\xi_{3}=\xi_{4}$ and setting $\Omega=\sqrt{\xi_{1}^{2}+\xi_{2}^{2}}$ we find that the solutions are given by,

\begin{eqnarray}
\hat{X}_{5}(t)&=&D\hat{X}_{5}(0)+E\hat{X}_{6}(0)+F\hat{X}_{7}(0)+G\hat{X}_{8}(0),\\
\hat{Y}_{5}(t)&=&D\hat{Y}_{5}(0)-E\hat{Y}_{6}(0)+F\hat{Y}_{7}(0)-G\hat{Y}_{8}(0),\\
\hat{X}_{6}(t)&=&E\hat{X}_{5}(0)+H\hat{X}_{6}(0)+E\hat{X}_{7}(0)+I\hat{X}_{8}(0),\\
\hat{Y}_{6}(t)&=&-E\hat{Y}_{5}(0)+H\hat{Y}_{6}(0)-E\hat{Y}_{7}(0)+I\hat{Y}_{8}(0),\\
\hat{X}_{7}(t)&=&F\hat{X}_{5}(0)+E\hat{X}_{6}(0)+D\hat{X}_{7}(0)+G\hat{X}_{8}(0),\\
\hat{Y}_{7}(t)&=&F\hat{Y}_{5}(0)-E\hat{Y}_{6}(0)+D\hat{Y}_{7}(0)-G\hat{Y}_{8}(0),\\
\hat{X}_{8}(t)&=&G\hat{X}_{5}(0)+I\hat{X}_{6}(0)+G\hat{X}_{7}(0)+J\hat{X}_{8}(0),\\
\hat{Y}_{8}(t)&=&-G\hat{Y}_{5}(0)+I\hat{Y}_{6}(0)-G\hat{Y}_{7}(0)+J\hat{Y}_{8}(0),
\end{eqnarray}

\noindent where

\begin{eqnarray}
D&=&\cosh^{2}(\Omega t/\sqrt{2}),\\
E&=&\sinh^{2}(\Omega t/\sqrt{2}),\\
F&=&\frac{\xi_{1}\sinh(\sqrt{2}\Omega t)}{\sqrt{2}\Omega},\\
G&=&\frac{\xi_{2}\sinh(\sqrt{2}\Omega t)}{\sqrt{2}\Omega},\\
H&=&\frac{\xi_{2}^{2}+\xi_{1}^{2}\cosh(\sqrt{2}\Omega t)}{\Omega^{2}},\\
I&=&\frac{\xi_{1}\xi_{2}(\cosh(\sqrt{2}\Omega t)-1)}{\Omega^{2}},\\
J&=&\frac{\xi_{1}^{2}+\xi_{2}^{2}\cosh(\sqrt{2}\Omega t)}{\Omega^{2}}.
\end{eqnarray}

\noindent We note here the generality of the solutions presented and that other cases are possible numerically. The variances are given by,

\begin{eqnarray}
\langle\hat{X}_{5}^{2}\rangle&=&\langle\hat{Y}_{5}^{2}\rangle=D^{2}+E^{2}+F^{2}+G^{2},\\
\langle\hat{X}_{6}^{2}\rangle&=&\langle\hat{Y}_{6}^{2}\rangle=2E^{2}+H^{2}+I^{2},\\
\langle\hat{X}_{7}^{2}\rangle&=&\langle\hat{Y}_{7}^{2}\rangle=D^{2}+E^{2}+F^{2}+G^{2},\\
\langle\hat{X}_{8}^{2}\rangle&=&\langle\hat{Y}_{8}^{2}\rangle=2G^{2}+I^{2}+J^{2},
\end{eqnarray}

\noindent  and the covariances are,

\begin{eqnarray}
\langle\hat{X}_{5}{X}_{6}\rangle&=&-\langle\hat{Y}_{5}{Y}_{6}\rangle=DE+EH+EF+IG,\\
\langle\hat{X}_{5}{X}_{7}\rangle&=&\langle\hat{Y}_{5}{Y}_{7}\rangle=2DF+E^{2}+G^{2},\\
\langle\hat{X}_{5}{X}_{8}\rangle&=&-\langle\hat{Y}_{5}{Y}_{8}\rangle=DG+EI+FG+GJ,\\
\langle\hat{X}_{6}{X}_{7}\rangle&=&-\langle\hat{Y}_{6}{Y}_{7}\rangle=DE+EH+EF+IG,\\
\langle\hat{X}_{6}{X}_{8}\rangle&=&\langle\hat{Y}_{6}{Y}_{8}\rangle=2EG+HI+IJ,\\
\langle\hat{X}_{7}{X}_{8}\rangle&=&-\langle\hat{Y}_{7}{Y}_{8}\rangle=DG+EI+FG+GJ.
\end{eqnarray}

\begin{figure}[ht!]
\includegraphics[width=0.85\columnwidth]{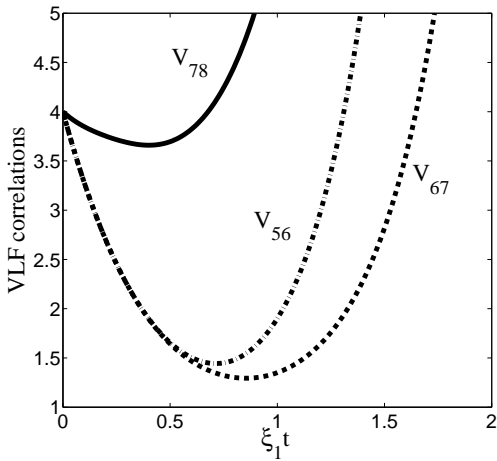} 
\caption{Analytic solutions for the optimized van Loock-Furusawa correlations, $V_{56}$, $V_{67}$ and $V_{78}$ with $\xi_{2}=0.5\xi_{1}$, found by solving the Heisenberg equations of motion in the undepleted pump approximation. Having all three of the correlations drop below 4 is sufficient to demonstrate quadripartite entanglement.}
\label{fig00}
\end{figure}

As described in Section~\ref{caseA}, we can now calculate the VLF correlations. Figure~\ref{fig00} shows the optimized VLF correlations as a function of $\xi_{1} t$, for the case $\xi_{2}=0.5\xi_{1}$. Again, we observe that quadripartite entanglement is present in this system with all three VLF correlations less than 4 for some $\xi_{1} t$. Comparing Figures~\ref{fig0} and~\ref{fig00} we see that the greatest degree of entanglement is obtained for the case where all $\xi_{i}$ are equal. Figure~\ref{fig00} shows that for each VLF correlation the entanglement is degraded beyond a particular value of $\xi_{1} t$, however this is not the case in Figure~\ref{fig0} when all the $\xi_{i}$ are equal.

\section{Equations of Motion for the full Hamiltonian}

We will now consider the full physical system, where the nonlinear media are contained inside a pumped resonant Fabry-P\'{e}rot cavity. The master equation for the density operator of the system can be found in the standard manner by tracing over the reservoirs \cite{Danbook} and is given by,

\begin{equation}
\frac{\partial \hat{\rho}}{\partial t} = -\frac{i}{\hbar}\Big[\hat{H}_{pump}+\hat{H}_{int}, \hat{\rho}\Big]+\sum_{i=1}^{8}\gamma_{i} {\cal{D}}_{i}[\hat{\rho}]
\label{master}
\end{equation}

\noindent where $\gamma_{i}$ are the cavity loss rates at the respective frequencies and $\cal{D}$$_{i}[\hat{\rho}]=2\hat{a}_{i} \hat{\rho}\hat{a}^{\dagger}_{i} -\hat{a}^{\dagger}_{i} \hat{a}_{i} \hat{\rho}-\hat{\rho}\hat{a}^{\dagger}_{i} \hat{a}_{i} $ is the Lindblad superoperator \cite{Danbook} under the usual zero-temperature Markov approximation. From this one can derive the stochastic differential equations (SDEs) in the positive-$P$ representation \cite{plusP}, and in turn, study the intracavity dynamics. 

Our approach \cite{GardinerQN} involves converting the quantum operator equations of motion of Equation~\ref{master} into a Fokker-Planck equation for the positive-$P$ representation pseudoprobability distribution of the system \cite{plusP, GardinerQN}. This can then be interpreted as a set of $c$-number SDEs. It should be noted that the use of the positive-$P$ representation, rather than the Glauber-Sudarshan P representation \cite{Glauber, Sudarshan}, is necessary to ensure that the diffusion matrix of the FPE is positive-definite. This is achieved with the positive-P approach by defining two independent stochastic fields $\alpha_{i}$ and $\alpha^{+}_{i}$ corresponding to the mode operators $\hat{a}_{i}$ and $\hat{a}^{\dagger}_{i}$, respectively, in the limit of a large number of stochastic trajectories. Using this method it is possible to calculate stochastic trajectory averages which correspond to the normally-ordered expectation values of quantum-mechanical operators, for example, $\overline{(\alpha_{i})^{n}(\alpha_{j}^{\tiny{+}})^{m}}=\langle (\hat{a}^{\dagger}_{j})^{m}\hat{a}_{i}^{n}\rangle$. Taking this approach yields a diffusion matrix of the form,

\begin{equation}
\boldsymbol{D}=\bordermatrix{&\cr &\bf{0}&\bf{0}\cr&\bf{0}&\boldsymbol{d}},
\end{equation}

\noindent where $\textbf{0}$ is an 8 $\times$ 8 null matrix and the non zero block is given by,

\begin{widetext}
\begin{equation}
\boldsymbol{d}=\bordermatrix{&\cr &0&0 &\chi_{1}\alpha_{1}&0 &0&0 &\chi_{4}\alpha_{4}&0\cr&0&0 &0&\chi_{1}\alpha_{1}^{\tiny{+}} &0&0 &0&\chi_{4}\alpha_{4}^{\tiny{+}} \cr&\chi_{1}\alpha_{1} &0 &0&0 &\chi_{2}\alpha_{2}&0 &0&0\cr&0&\chi_{1}\alpha_{1}^{\tiny{+}} &0&0 &0&\chi_{2}\alpha_{2}^{\tiny{+}} &0&0\cr&0&0 &\chi_{2}\alpha_{2}&0 &0&0 &\chi_{3}\alpha_{3}&0\cr&0&0 &0&\chi_{2}\alpha_{2}^{\tiny{+}}  &0&0 &0&\chi_{3}\alpha_{3}^{\tiny{+}} \cr&\chi_{4}\alpha_{4}&0 &0&0 &\chi_{3}\alpha_{3}&0 &0&0\cr&0&\chi_{4}\alpha_{4}^{\tiny{+}} &0&0 &0&\chi_{3}\alpha_{3}^{\tiny{+}} &0&0}.
\end{equation}
\end{widetext}

\noindent The matrix $\boldsymbol{d}$ can be factorised such that the It\^o stochastic differential equations are obtained. For the high frequency fields this process yields,

\begin{eqnarray}
\frac{d\alpha_{1}}{dt}&=&\epsilon_{1}-\chi_{1}\alpha_{5}\alpha_{6}-\gamma_{1}\alpha_{1},\nonumber\\
\frac{d\alpha_{2}}{dt}&= & \epsilon_{2}-\chi_{2}\alpha_{6}\alpha_{7}-\gamma_{2}\alpha_{2},\nonumber\\
\frac{d\alpha_{3}}{dt}&=& \epsilon_{3}-\chi_{3}\alpha_{7}\alpha_{8}-\gamma_{3}\alpha_{3},\nonumber\\
\frac{d\alpha_{4}}{dt}&=&\epsilon_{4}-\chi_{4}\alpha_{8}\alpha_{5}-\gamma_{4}\alpha_{4},\nonumber\\
\label{posp1}
\end{eqnarray}

\noindent and also the equations found by interchanging $\alpha_{i}$ and $\alpha_{i}^{+}$. Whilst for the low frequency fields one obtains,

\begin{eqnarray}
\frac{d\alpha_{5}}{dt}&=&\chi_{1}\alpha_{1}\alpha_{6}^{\tiny{+}}+\chi_{4}\alpha_{4}\alpha_{8}^{\tiny{+}}-\gamma_{5}\alpha_{5}+\sqrt{\frac{\chi_{1}\alpha_{1}}{2}}(\eta_{5}(t)+i\eta_{6}(t))\nonumber\\&+&\sqrt{\frac{\chi_{4}\alpha_{4}}{2}}(\eta_{13}(t)+i\eta_{14}(t)),\nonumber\\
\frac{d\alpha_{6}}{dt}&=& \chi_{1}\alpha_{1}\alpha_{5}^{\tiny{+}}+\chi_{2}\alpha_{2}\alpha_{7}^{\tiny{+}}-\gamma_{6}\alpha_{6}+\sqrt{\frac{\chi_{1}\alpha_{1}}{2}}(\eta_{5}(t)-i\eta_{6}(t))\nonumber\\&+&\sqrt{\frac{\chi_{2}\alpha_{2}}{2}}(\eta_{9}(t)+i\eta_{10}(t)),\nonumber\\
\frac{d\alpha_{7}}{dt}&=& \chi_{2}\alpha_{2}\alpha_{6}^{\tiny{+}}+\chi_{3}\alpha_{3}\alpha_{8}^{\tiny{+}}-\gamma_{7}\alpha_{7}+\sqrt{\frac{\chi_{3}\alpha_{3}}{2}}(\eta_{1}(t)+i\eta_{2}(t))\nonumber\\&+&\sqrt{\frac{\chi_{2}\alpha_{2}}{2}}(\eta_{9}(t)-i\eta_{10}(t)),\nonumber\\
\frac{d\alpha_{8}}{dt}&=&\nonumber  \chi_{3}\alpha_{3}\alpha_{7}^{\tiny{+}}+\chi_{4}\alpha_{4}\alpha_{5}^{\tiny{+}}-\gamma_{8}\alpha_{8}+\sqrt{\frac{\chi_{3}\alpha_{3}}{3}}(\eta_{1}(t)-i\eta_{2}(t))\nonumber\\&+&\sqrt{\frac{\chi_{4}\alpha_{4}}{2}}(\eta_{13}(t)-i\eta_{14}(t)),\nonumber\\
\label{posp}
\end{eqnarray}

\noindent and also the equations found by interchanging $\alpha_{i}$ and $\alpha_{i}^{+}$ and $\eta_{i}(t)$ and $\eta_{i+2}(t)$. The $\gamma_{i}$ are the cavity loss rates at the respective frequencies, and $\eta_{i}(t)$ are real, independent, Gaussian noise terms which satisfy $\overline{\eta_{i}(t)}=0$ and $\overline{\eta_{i}(t)\eta_{j}(t^{\prime})}=\delta_{ij}\delta(t-t^{\prime})$. It should be mentioned that we are assuming that all the intracavity modes are resonant with the cavity, and although it is possible to include detuning we do not do so here. 

\section{Stability Analysis and Fluctuation Spectra}

We conduct a linearized fluctuation analysis \cite{Danbook} of the system for the purposes of calculating the output spectral correlations for the cavity from the intracavity spectra. We begin by neglecting the noise terms in Equation~(\ref{posp}) so that $\alpha_{i}^{+} \rightarrow \alpha_{i}^{*}$, and also set $\alpha_{i}=\bar{\alpha}_{i}+\delta\alpha_{i}$, where $\bar{\alpha}_{i}$ is a mean value and $\delta\alpha_{i}$ represents the fluctuations. This gives a set of classical equations for the mean values,

\begin{eqnarray}
\frac{d\bar\alpha_{1}}{dt}&=&\epsilon_{1}-\chi_{1}\bar\alpha_{5}\bar\alpha_{6}-\gamma_{1}\bar\alpha_{1},\nonumber\\
\frac{d\bar\alpha_{2}}{dt}&= & \epsilon_{2}-\chi_{2}\bar\alpha_{6}\bar\alpha_{7}-\gamma_{2}\bar\alpha_{2},\nonumber\\
\frac{d\bar\alpha_{3}}{dt}&=& \epsilon_{3}-\chi_{3}\bar\alpha_{7}\bar\alpha_{8}-\gamma_{3}\bar\alpha_{3},\nonumber\\
\frac{d\bar\alpha_{4}}{dt}&=&\epsilon_{4}-\chi_{4}\bar\alpha_{8}\bar\alpha_{5}-\gamma_{4}\bar\alpha_{4}\nonumber\\
\frac{d\bar\alpha_{5}}{dt}&=&\chi_{1}\bar\alpha_{1}\bar\alpha_{6}^{*}+\chi_{4}\bar\alpha_{4}\bar\alpha_{8}^{*}-\gamma_{5}\bar\alpha_{5},\nonumber\\
\frac{d\bar\alpha_{6}}{dt}&=& \chi_{1}\bar\alpha_{1}\bar\alpha_{5}^{*}+\chi_{2}\bar\alpha_{2}\bar\alpha_{7}^{*}-\gamma_{6}\bar\alpha_{6},\nonumber\\
\frac{d\bar\alpha_{7}}{dt}&=& \chi_{2}\bar\alpha_{2}\bar\alpha_{6}^{*}+\chi_{3}\bar\alpha_{3}\bar\alpha_{8}^{*}-\gamma_{7}\bar\alpha_{7},\nonumber\\
\frac{d\bar\alpha_{8}}{dt}&=&\nonumber  \chi_{3}\bar\alpha_{3}\bar\alpha_{7}^{*}+\chi_{4}\bar\alpha_{4}\bar\alpha_{5}^{*}-\gamma_{8}\bar\alpha_{8},\\
\label{posp2}
\end{eqnarray}
and from these we can obtain steady-state solutions. 

In the remainder of this paper we consider a symmetric system where all the high frequency modes have the same cavity damping rates, with $\gamma_{i}=\gamma$ for $i=1,2,3,4$ and all low frequency modes also have equal cavity damping rates, with $\gamma_{i}=\kappa$ for $i=5,6,7,8$. In addition to this, we assume all the nonlinearities and hence all the pump field amplitudes are equal, that is, $\chi_{i}=\chi$ and $\epsilon_{i}=\epsilon$, respectively.

For this completely symmetric system, we verify that there is an oscillation threshold at the critical pumping amplitude,
 
\begin{equation}
\epsilon_{c}=\frac{\gamma\kappa}{2\chi}
\end{equation}

\noindent as is the case for triply concurrent downconversion \cite{mko3}. This result differs from the standard non-degenerate OPO threshold condition by a factor of a half. This difference here is due to the fact that each pump mode drives two down-conversion processes. The stationary solutions below this threshold value are found to be, 

\begin{eqnarray}
\bar{\alpha}_{i}=\frac{\epsilon}{\gamma} \hspace{1cm} \textnormal{for} \hspace{0.1cm}i \in \{1,2,3,4\},\nonumber\\
\bar{\alpha}_{i}=0 \hspace{1cm} \textnormal{for} \hspace{0.1cm}i \in \{5,6,7,8\}.
\end{eqnarray}

\noindent Whilst, above threshold the stationary solutions are given by, 

\begin{equation}
\bar{\alpha}_{i}=\frac{\kappa}{2\chi} \hspace{1cm} \textnormal{for} \hspace{0.1cm}i \in \{1,2,3,4\}
\end{equation}

\begin{equation}
\bar{\alpha}_{i}=\sqrt{(\epsilon-\epsilon_{c})/\chi }\hspace{1cm} \textnormal{for} \hspace{0.1cm}i \in \{5,6,7,8\}.
\end{equation}

We see that the low frequency modes become macroscopically occupied as the pumping is increased and the high frequency modes remain at their threshold value. Using Equations~\ref{posp1} and~\ref{posp} we also perform dynamical simulations to confirm the steady state values. 

We then proceed to study fluctuations around the steady state which allows one to calculate measurable output fluctuation spectra \cite{GardinerQN} and hence, quantify the quantum correlations of the system. The linearized equations for the fluctuations are of the form,

\begin{equation}
d\boldsymbol{\delta\alpha}=-\boldsymbol{\bar{A}\delta\alpha} dt + \boldsymbol{\bar{B}} d\boldsymbol{W},
\end{equation}

\noindent where $\boldsymbol{\delta\alpha}=[\delta\alpha_{1}, \delta\alpha^{+}_{1},\delta\alpha_{2},\delta\alpha^{+}_{2},\dots,\delta\alpha_{8},\delta\alpha^{+}_{8}]^{T}$, $\bf{\bar B}$ is the noise matrix of Equation~\ref{posp} with the steady-state values inserted, $d\boldsymbol{W}$ is a vector of independent, real Wiener increments \cite{GardinerQN} and $\boldsymbol{\bar A}$ is the drift matrix with the steady-state values inserted and given by,

\begin{equation}
	 \boldsymbol{\bar{A}}=\bordermatrix{ & \cr & \boldsymbol{A}_{1} & \boldsymbol{A}_{2} \cr & \cr & -(\boldsymbol{A}_{2}^{*})^{T} & \boldsymbol{A}_{3} \cr  },
	\label{sarah}
\end{equation}

\noindent where $\boldsymbol{A}_{1}=-\gamma I_{6}$,

\begin{widetext}
\begin{equation}
	 \boldsymbol{A}_{2}=\bordermatrix{ & \cr & -\chi_{1}\bar\alpha_{6} & 0  & -\chi_{1}\bar\alpha_{5} & 0  & 0 & 0 & 0  & 0\cr  & 0  & -\chi_{1}\bar\alpha_{6}^{*}  & 0  & -\chi_{1}\bar\alpha_{5}^{*} & 0& 0  & 0 & 0 & \cr  & 0 & 0 &  -\chi_{2}\bar\alpha_{7}  & 0  & -\chi_{2}\bar\alpha_{6} & 0 &0  & 0&\cr   &0  & 0 &0  & -\chi_{2}\bar\alpha_{7}^{*} & 0 & -\chi_{2}\bar\alpha_{6}^{*} &0&0& \cr  & 0 & 0 & 0 & 0 & -\chi_{3}\bar\alpha_{8}&0  & -\chi_{3}\bar\alpha_{7} & 0 & \cr  & 0 & 0  & 0  & 0 & 0  & -\chi_{3}\bar\alpha_{8}^{*} & 0  & -\chi_{3}\bar\alpha_{7}^{*} & \cr & -\chi_{4}\bar\alpha_{8} & 0 & 0  & 0 & 0 & 0  & -\chi_{4}\bar\alpha_{5} &0 &\cr& 0  & -\chi_{4}\bar\alpha_{8}^{*} & 0  & 0 & 0 & 0  & 0& -\chi_{4}\bar\alpha_{5}^{*}},
	\label{rr}
\end{equation}

\noindent and

\begin{equation}
	 \boldsymbol{A}_{3}=\bordermatrix{ & \cr & -\kappa & 0  & 0 & \chi_{1}\bar\alpha_{1} & 0 & 0 & 0  & \chi_{4}\bar\alpha_{4} \cr  & 0  &  -\kappa  & \chi_{2}\bar\alpha_{1}^{*}  & 0 & 0 & 0&  \chi_{4}\bar\alpha_{4}^{*} & 0\cr  & 0 & \chi_{1}\bar\alpha_{1}  &  -\kappa  & 0  & 0 & \chi_{2}\bar\alpha_{2} & 0 & 0\cr  & \chi_{1}\bar\alpha_{1}^{*}  & 0& 0 & -\kappa  & \chi_{2}\bar\alpha_{2}^{*}   &  0 & 0 & 0 & \cr  & 0 & 0  & 0 & \chi_{2}\bar\alpha_{2}  &  -\kappa& 0 & 0 & \chi_{3}\bar\alpha_{3}\cr  & 0 & 0& \chi_{2}\bar\alpha_{2}^{*}   & 0 & 0  &  -\kappa &\chi_{3}\bar\alpha_{3}^{*} & 0& \cr & 0  & \chi_{4}\bar\alpha_{4} & 0  & 0 & 0 & \chi_{3}\bar\alpha_{3} &  -\kappa &0 &\cr& \chi_{4}\bar\alpha_{4}^{*}  & 0 & 0  & 0 &  \chi_{3}\bar\alpha_{3}^{*} & 0  & 0&  -\kappa}.
	\label{yy}
\end{equation}

\end{widetext}

\noindent Provided the eigenvalues of the drift matrix $\boldsymbol{\bar A}$ have no negative real part the system is stable and we can treat the fluctuation equations as describing an Ornstein-Uhlenbeck process \cite{SMCrispin}. This allows one to calculate the intracavity spectral correlation matrix,

\begin{equation}
\boldsymbol{S}(\omega)=(\boldsymbol{\bar{A}}+i\omega \openone)\boldsymbol{\bar{B}\bar{B}}^{T}(\boldsymbol{\bar{A}}^{T}-i\omega \openone)^{-1},
\end{equation}

\noindent and this is related to the measurable output fluctuation spectra using the standard input-output relations for optical cavities \cite{collett}. Furthermore, it supplies us with all that is necessary to calculate the measurable extracavity quadripartite entanglement.

\section{OUTPUT FLUCTUATION SPECTRA}

\begin{figure}[ht!]
\includegraphics[width=0.95\columnwidth]{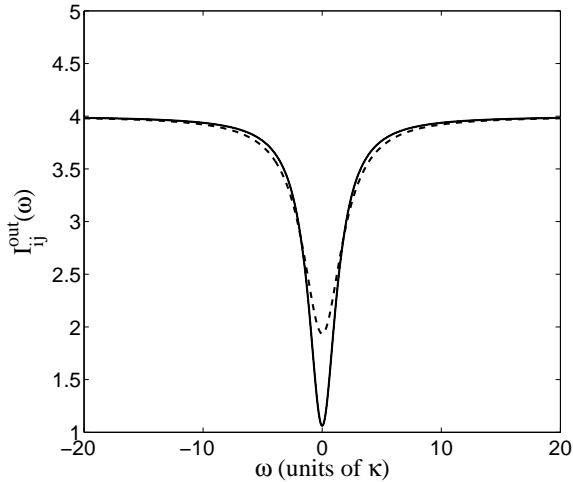} 
\caption{The output spectral correlations,  $I^{out}_{ij}(\omega)$, as a function of frequency $\omega$ (units of $\kappa$) corresponding to the quadripartite entanglement criteria given in Equation 21 - 23. The below threshold (solid line) and above threshold (dashed line) cases are shown, for cavity pump amplitudes $\epsilon=0.8\epsilon_{c}$ and $\epsilon=1.2\epsilon_{c}$, respectively. The three correlations are equal for each case for the chosen parameters, which are symmetric with $\chi_{i}=\chi$, $\epsilon_{i}=\epsilon$, $\gamma=10$, $\kappa=1$ and $\chi=10^{-2}$. All quantities plotted here and in the subsequent graph are dimensionless.}
\label{fig1}
\end{figure}

The same inequalities as given in Section~\ref{III} in terms of variances also hold when expressed in terms of the output spectra, and it is these quantities that can be measured in experiments. In the following, we use the notation  $I^{out}_{ij}(\omega)$ (ie. any of $I^{out}_{56}, I^{out}_{67}, I^{out}_{78}$) to represent the three output spectral correlations of the same form as the optimized expressions in Eqns. (\ref{V56}) - (\ref{V78}). In Figure~\ref{fig1}, we plot these three correlations as a function of frequency for our completely symmetric system for the below threshold case (solid line) and the above threshold case (dashed line). In both cases, the three correlations are equal, ie. $I^{out}_{56}=I^{out}_{67}= I^{out}_{78}$. 

The inequalities are violated both below and above the threshold, demonstrating quadripartite entanglement. Specifically, the results shown are for the case of the pump field amplitudes set at $\epsilon=0.8\epsilon_{c}$ and $\epsilon=1.2\epsilon_{c}$. For these parameter choices, the largest violation of the van Loock-Furusawa entanglement criteria, and thus the maximum quadripartite entanglement, is for the low frequency modes below threshold. In general, for both cases the largest degree of violation of the inequalities is observed near zero frequency. For large frequencies $I^{out}_{ij}(\omega) \rightarrow 4$ which is the uncorrelated limit for our optimized expressions.  


\begin{figure}[ht!]
\includegraphics[width=0.95\columnwidth]{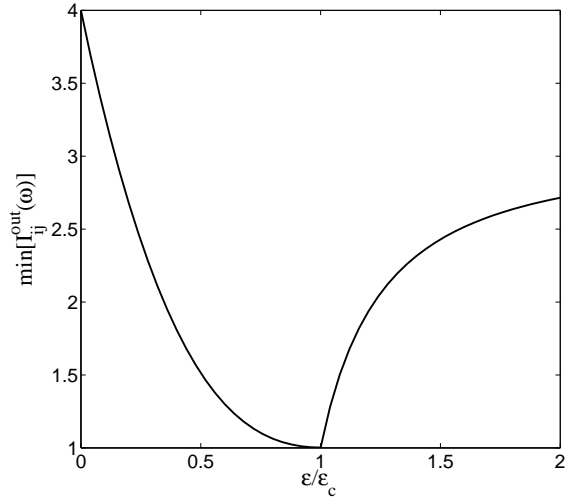}
\caption{Maximum quadripartite entanglement as a function of the ratio of the cavity pumping to the pumping threshold. The cavity parameters are the same as in Figure~\ref{fig1}, with $\gamma=10$, $\kappa=1$ and $\chi=10^{-2}$. Again, all three correlations are equal for these parameters. It should be stressed that at $\epsilon/\epsilon_{c} = 1$ the validity of the results is limited as the linearized analysis is no longer valid.}
\label{fig2}
\end{figure}

We also determine the maximum quadripartite entanglement for the same parameters as in Figure~\ref{fig1}, but for a range of pump field amplitudes on both sides of the oscillation threshold. This is shown in Figure~\ref{fig2}, where we plot the minimum value of the output spectra at any frequency, as a function of $\epsilon/\epsilon_{c}$. As expected \cite{muzzy1,muzzy}, we observe the maximum quadripartite entanglement at the critical pumping amplitude, with the caveat that the linearized fluctuation analysis is not valid in the immediate vicinity of the threshold. The fact that a gradual slope is observed below threshold in the region of maximum entanglement could prove useful for future experimental realizations of this scheme. It is also found that quadripartite entanglement persists well above threshold, with a large violation of the van Loock-Furusawa criteria still present as the pumping is increased above the critical pumping value. In particular, well above threshold the minimum of $I^{out}_{ij}(\omega)$ approaches 3. This behaviour is also seen in \cite{muzzy1,muzzy} where the equivalent correlations also asymptote to a finite value.

\section{Conclusions and outlook}
We have demonstrated intracavity continous-variable quadripartite entanglement in quadruply concurrent downconversion, both above and below the critical pumping threshold using optimized van Loock-Furusawa criteria. Above threshold the proposed scheme produces a source of bright entangled output beams. The below threshold regime provides the greatest degree of entanglement and a region where this entanglement could be measured in experiments. One of the advantages of this type of scheme lies in the number of different regimes that can be explored by tuning various parameters in experiments. For example, the pump intensities and coupling strengths can be tuned and this makes it possible to vary the degree of entanglement in the system.

Throughout this article, we have studied the properties of the interaction Hamiltonian, presented the full quantum equations of motion and performed a linearized fluctuation analysis. All results indicate that this system is a good candidate for the demonstration of quadripartite continuous-variable entanglement. In relation to experimental implementation of the scheme presented here, stabilizing a single cavity in which four entangled modes are created may prove preferable in some applications rather than alternative schemes which rely on stabilizing and synchronizing multiple OPOs. 

Finally, this result could be of further significance in a similar system where one of the nonlinear couplings is absent. Such a system may be a candidate for realizing the simplest four node cluster state \cite{zaidi, men2}.

\section{Acknowledgments}
SLWM and MKO are supported by the Australian Research Council Centre of Excellence for Quantum-Atom Optics. SLWM would also like to thank the AFUW for the provision of a Georgina Sweet Fellowship. ASB is supported by the New Zealand Foundation for Research, Science, and Technology under Contract No. UOOX0801. 


\end{document}